\documentclass[a4paper,11pt]{article}
\usepackage{jinstpub} % for details on the use of the package, please see the JINST-author-manual
\usepackage{lineno}
\usepackage{siunitx}
\usepackage{xspace} % to keep a space after the micrometer without having to type {}
\usepackage[utf8]{inputenc}
\usepackage[T1]{fontenc}
\def\um{\si{\micro\m}\xspace}

\title{\boldmath A data acquisition and reconstruction software for SAMPA-SRS integration.}

\author[a,1]{G. G. A. de Souza,\note{Corresponding author.}}
\author[a]{T. S. Abelha,}
\author[a]{T. B. Saramela,}
\author[b]{A. F. V. Cortez,}
\author[b]{H. N. da Luz,}
\author[a]{C. G. Penteado,}
\author[a]{M. Bregant,}

\affiliation[a]{Instituto de Física da Universidade de São Paulo\\
Rua do Matão 1371, 05508-090 Cidade Universitária, São Paulo, Brasil}
\affiliation[b]{Institute of Experimental and Applied Physics, Czech Technical University in Prague\\ Husova 5, 110 00 Prague 1, Czech Republic}

\emailAdd{geovane.souza@usp.br}

\abstract{
In this work we present the latest developments in the SAMPA-SRS integration. A
software was developed to improve the acquisition configuration, acquisition, and decoding of the data. The complete framework was tested using a triple GEM-based position sensitive detector for X-rays. The detector was operated in Ar/CO$_2$ (70/30) in continuous flow, at atmospheric pressure and made use of a 1 dimension strip readout (\SI{200}{\micro\meter} wide strips at a pitch of \SI{400}{\micro\meter}) for charge collection . With this detector a position resolution of better than \SI{833}{\micro\meter} was obtained, with an energy resolution of 14.2\,\%\,($\sigma/E$) for \SI{5.9}{\kilo\electronvolt}.
}


\begin{document}
\maketitle
\flushbottom

\section{Introduction}

%Either by looking for rare events that require experiments working at a high count rate, or by improve the imaging quality capability of gaseous detectors, there is a requirement to improve the acquisition electronic systems. 
The Scalable Readout System~(SRS)~\cite{SRS} is an acquisition system that works as a common back-end electronics to collect and transmit data from different detectors as a complete and scalable system. 
Different ASICS, such as the APV25~\cite{apv} or the VMM~\cite{vmm}, have already been successfully integrated and operated with this system. The lack of an alternative for TPC readout electronics capable of working in continuous mode, in systems that can range from a few hundreds up to thousands of channels, was a motivation for the incorporation of the SAMPA ASIC, developed for ALICE's TPC and Muon Chambers~\cite{Sampa}, in the SRS. The SAMPA can record waveforms of up to \SI{100}{\micro\second}, in time bins of 100\,ns, making the system suitable for reading the full active volume of Time Projection Chambers~(TPC) of practically any size. 
The SAMPA chip is installed in hybrid boards that are directly connected to the segmented readout electrodes of the detector. Each hybrid board contains 4 chips, corresponding to 128 electronic channels (32 channels per chip). 
The SAMPA-SRS integration has been tested in a small scale TPC ~\cite{SampaTPC}, successfully reconstructing cosmic tracks. Since then, both the firmware and the software tools have undergone important developments that resulted in improved performance in terms of data integrity, convenience of operation and data quality monitoring. In this work we summarize the latest updates on the development of the SAMPA-SRS controller and acquisition software. Some results obtained with an X-ray imaging detector using this system will be shown.

\section{The SampaSRS software}

A software package was developed to operate the SAMPA-SRS integration. Each module of the software is dedicated to a different part of the operation. The different steps of the framework were written in C++, using TCPDUMP \& LIBPCAP libraries for network traffic capture. CERN's ROOT~\cite{root} libraries were also implemented for data handling and storage.
The first module of the software (\emph{sampa\_control}) is in charge of the communication between the computer and the SAMPA chips (using I$^2$C protocol) to configure the acquisition (start/stop, internal/external trigger, waveform length, latency, zero suppression threshold).

A second separated tool (\emph{sampa\_acquisition}) captures the data packets from the network. If the user wants to have a real time preview of the acquisition, a graphic user interface (\emph{sampa\_gui}) processes and decodes small samples of the data to create control histograms (energy spectrum, position of the strip with the maximum energy and its waveform).
The \emph{sampa\_decoder} is used to transform the raw data produced by SAMPA into a root tree. The built events contain information of time stamp, channel and the complete waveform produced after a trigger. 

To operate in both zero or non zero suppression mode, it is necessary to produce and processes a pedestal file containing information of the baseline level and noise of each channel. It can be used for offline zero suppression or as input for the online zero suppression of SAMPA (the operation of the SAMPA chip in the zero suppression mode is implemented in the electronics, but it still not complete in this software package). The software also provides a simple clustering algorithm, that will be described in detail in the next section. This package processes the waveforms and generates a new tree with information of the center of mass of the cluster, its energy, its size, the time information and other relevant parameters. 

\section{Detector setup}

\begin{figure}[htbp]
\centering % \begin{center}/\end{center} takes some additional vertical space
\includegraphics[width=0.30\textwidth]{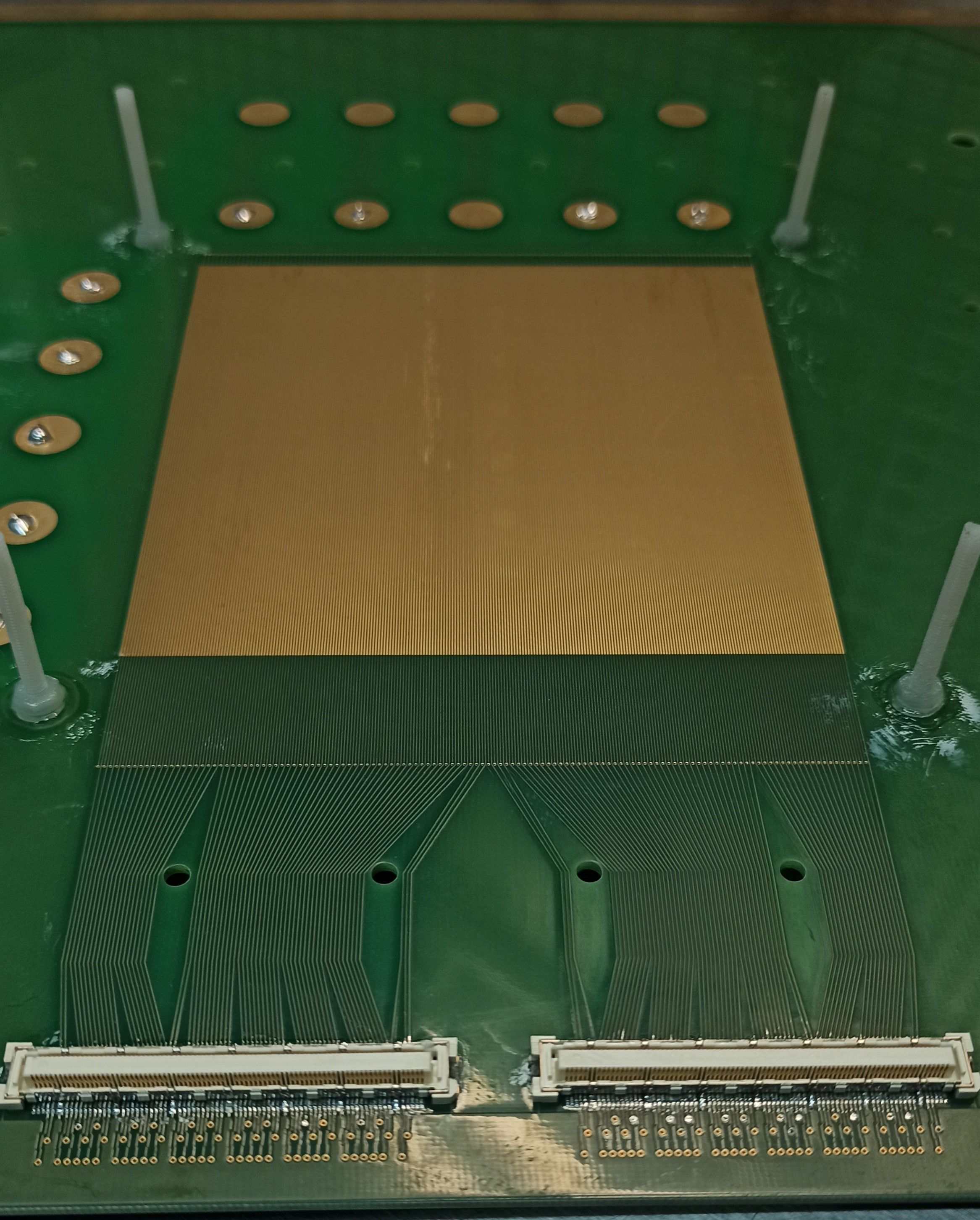}
\qquad
\includegraphics[width=.40\textwidth]{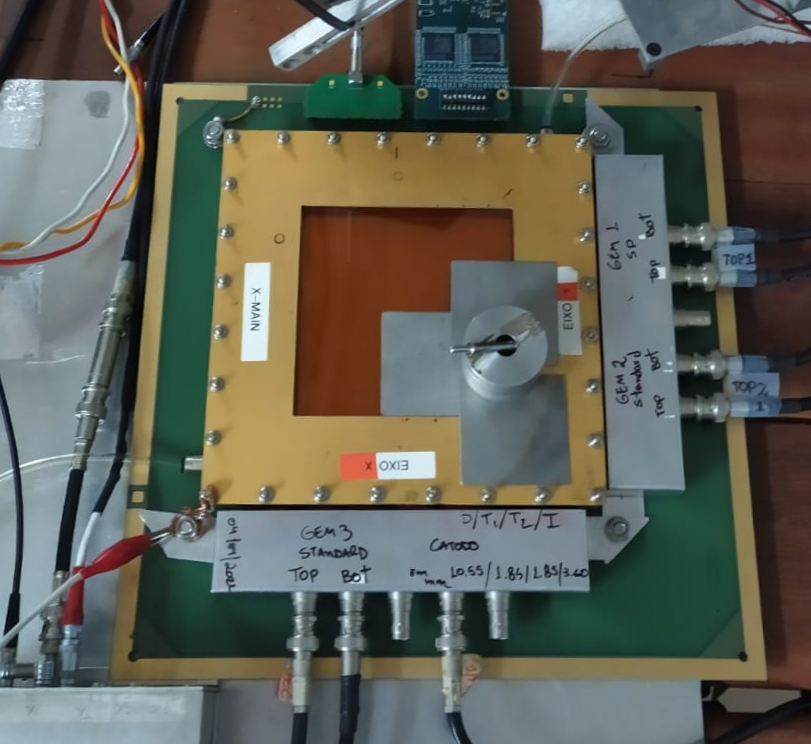}
% "\includegraphics" from the "graphicx" permits to crop (trim+clip)
% and rotate (angle) and image (and much more)
\caption{\label{fig:detector} Left: the 1D readout used in this work. It is composed of 256 independent readout strips with a width of \SI{200}{\micro\meter} at a pitch of \SI{400}{\micro\meter}. This detector is designed to accommodate two SAMPA hybrids. Right: The experimental setup. A $^{55}$Fe radioactive source was used to generate the signals in the detector for energy resolution studies (for the position resolution, a commercial X-ray tube was used). One SAMPA hybrid can be seen at the top of the photograph. }
\end{figure}

A hybrid board, equipped with four SAMPA chips, was mounted in an X-ray, 1D position sensitive detector based on a standard triple-GEM stack, with an absorption region of \SI{10.5}{\milli\meter} and operating in Ar/CO$_2$(70/30) at atmospheric pressure in open flow of 6 L/h. Figure~\ref{fig:detector} shows the segmented strip readout used for the charge collection. It consists of 256 strips, \SI{200}{\micro\meter} wide with a pitch of \SI{400}{\micro\meter}, spanning the complete $10\times 10$\,\si{\square\cm} sensitive area of the GEM stack, suited for the readout by two SAMPA hybrids. 
A resolution mask (standard Type 38 with 50\,\um-thick Pb layer) was placed on the detector window and a commercial X-ray tube (Amptek mini-X with Ag target) was used to to irradiate the detector with accelerating potential of \SI{10}{\kilo\volt}. The tube was placed \SI{30}{\centi\meter} away from the mask, for the purpose of testing the performance of the SAMPA chip and the acquisition software.

\section{Data processing and results}

To reduce the effect of the electronic noise on the data reconstruction, an offline zero suppression was applied. A lower level threshold around 4 times above the noise's $\sigma$ was used to distinguish the signals from the noise. The mean noise calculated over all strips was $\sigma=5.9(2)$ ADC channels. This value can be further improved by adjusting ground connections which are not optimal and by applying a common-mode correction to the data. The clustering algorithm starts to compose a cluster when a strip with a signal above the $4\sigma$ threshold is found. The neighboring strips are added to that cluster if they are above threshold until two consecutive strips are found below it.

\begin{figure}[htbp]
\centering % \begin{center}/\end{center} takes some additional vertical space
\includegraphics[width=.55\textwidth]{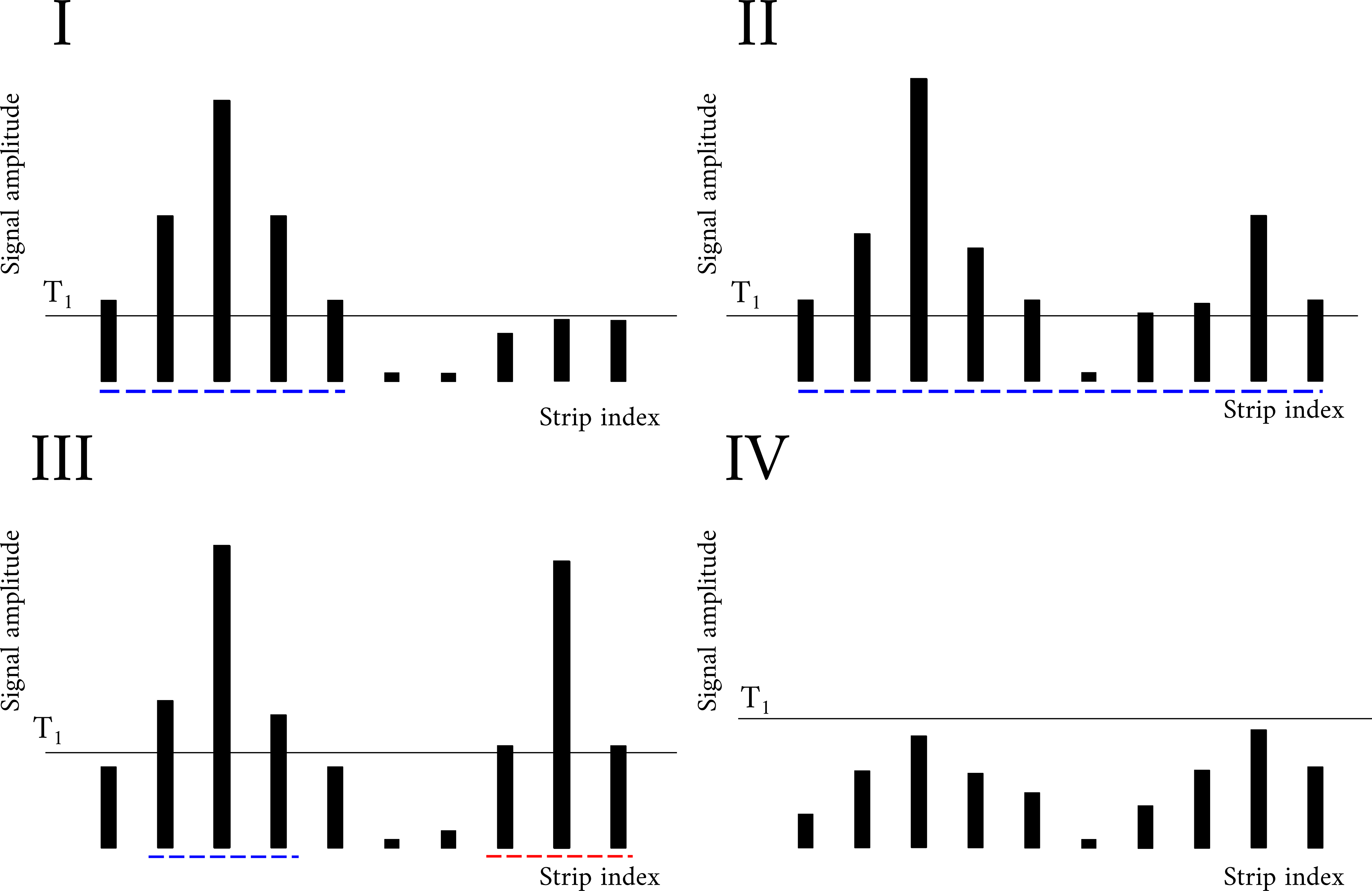}
\caption{\label{fig:threshold} Different examples of the result of the clustering algorithm. The valid clusters are marked with dashed lines\,(blue and red) in the x-axis. I) Valid cluster in the left. II) Valid cluster with no cluster split. III) Two clusters are created and they are both valid~(above T$_1$). IV) No valid cluster~(all signals below T$_1$)}
\end{figure}

Looking at figure \ref{fig:threshold}, it is possible to notice that a cluster consists of consecutive strips where the charge is collected at a greater value than the threshold (T$_1$). If two or more consecutive strips have value below T$_1$ the reconstruction of the current cluster is closed and another one can start.

\begin{figure}[htbp]
\centering % \begin{center}/\end{center} takes some additional vertical space
\includegraphics[width=.35\textwidth]{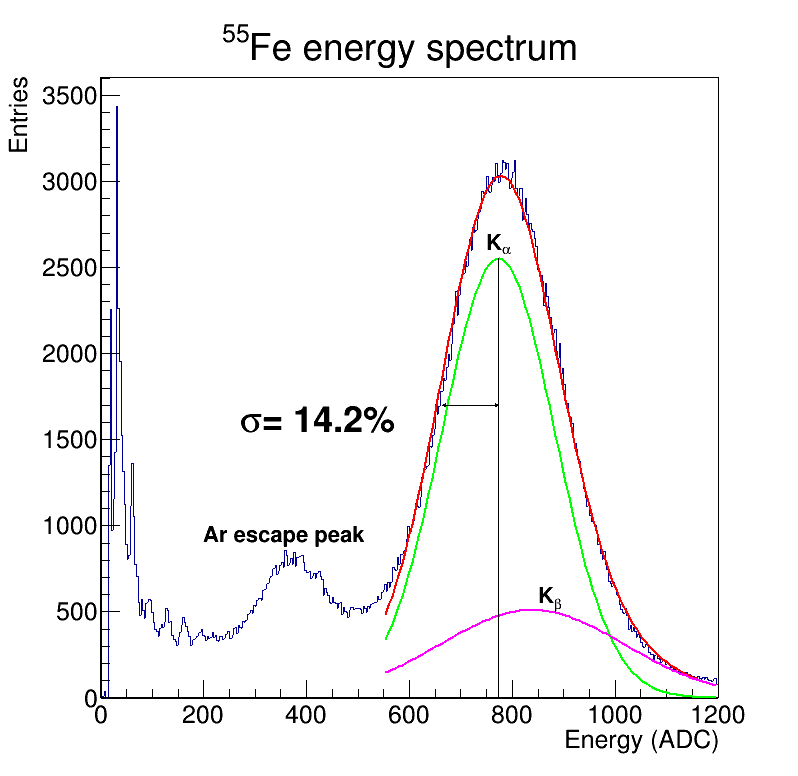} \includegraphics[width=.35\textwidth]{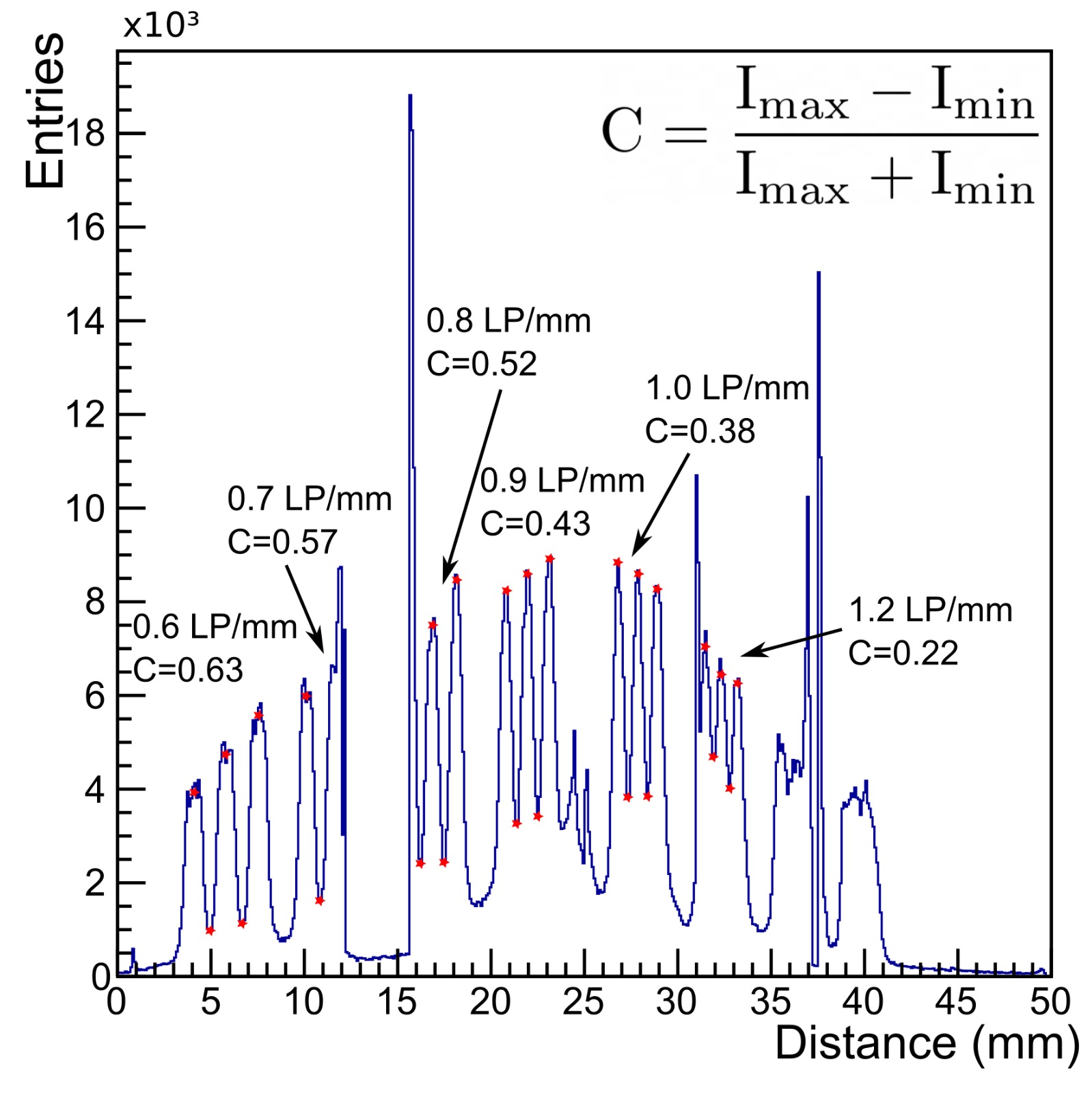}
\caption{\label{fig:resolution} Left: Energy spectrum. Right: Spatial resolution and contrast calculated using a lead mask and a commercial X-ray tube as source.}
\end{figure}

Figure~\ref{fig:resolution}, on the left, shows the pulse height distribution obtained with a $^{55}$Fe radioactive source, using the SAMPA chip. For this, the maximum signal amplitude of each strip within a cluster was considered. The energy resolution ($\sigma/E$) obtained was  14.2\% for \SI{5.9}{\kilo\electronvolt}. The distribution on the right is the 1D profile of the resolution mask, mentioned in the previous section, that allows to estimate the contrast at different spatial frequencies\,(the Contrast Transfer Function). The artefacts seen in the image are generated by dead channels in the SAMPA chip that led to floating strips, resulting in the electrons being pulled towards the closest working channel. The slit groups where the profile is corrupted by these artefacts were not used to calculate the contrast. The red dots represent the points used to calculate the contrast values for the different spatial frequencies and the position resolution was settled to be equivalent to the contrast at 10\%~\cite{medicalImage}. In this case better than \SI{833}{\micro\meter}~(1.2LP/mm). 

\section{Conclusion and future work}
The SAMPA chip has been tested, integrated in CERN/RD51 Scalable Readout System, this time in a thin-gap detector, only with position and energy resolution (no timing information), using  improved firmware and software tools. It is currently working with a negligible data loss. A complete tool for control, acquisition and data processing was developed and the reconstruction is made with small effort. It has been shown that although SAMPA was first developed for TPC applications, it may also work for 2D X-ray imaging. In the future we plan to explore the data acquisition with higher counting rates in zero suppression mode, recording only the maximum value of the ADC.

\acknowledgments
This study was financed in part by the Coordenação de Aperfeiçoamento de Pessoal de Nível Superior – Brasil (CAPES) – Finance Code 001,  by the grant 2014/12664-3 from Fundação de Amparo à Pesquisa do Estado de São Paulo (FAPESP). H.N. da Luz and A.F.V. Cortez acknowledge grant GAČR GA21-21801S (Czech Science Foundation).

% Bibliography
%% [A] Using JHEP.bst file
\bibliographystyle{JHEP}
\bibliography{biblio.bib}

\providecommand{\href}[2]{#2}\begingroup\raggedright\begin{thebibliography}{1}

\bibitem{SRS}
S.~Martoiu
  et~al.\href{https://doi.org/10.1088/1748-0221/8/03/c03015}{\emph{JINST}
  {\bfseries 8} (2013) C03015}.

\bibitem{apv}
M.J.~French
  et~al.\href{https://doi.org/10.1016/S0168-9002(01)00589-7}{\emph{Nucl. Instr.
  Meth. A} {\bfseries A466} (2001) 359}.

\bibitem{vmm}
G.~Iakovidis\href{https://doi.org/10.1088/1742-6596/1498/1/012051}{\emph{Journal
  of Physics: Conference Series} {\bfseries 1498} (2020) 012051}.

\bibitem{Sampa}
H.~Hernández et~al.\href{https://doi.org/10.1109/TIM.2019.2931016}{\emph{IEEE
  Transactions on Instrumentation and Measurement} {\bfseries 69} (2020) 2686}.

\bibitem{SampaTPC}
G.G.A.~{de Souza}, A.F.~Cortez, C.G.~Penteado, H.N.~{da Luz} and
  M.~Bregant\href{https://doi.org/https://doi.org/10.1016/j.nima.2022.167577}{\emph{Nucl.
  Instr. Meth. A} {\bfseries 1045} (2023) 167577}.

\bibitem{root}
R.~Brun and F.~Rademakers{\emph{Nucl. Instr. Meth. A} {\bfseries 389} (1997)
  91}.

\bibitem{medicalImage}
W.R.~Hendee and E.R.~Ritenour, Wiley, New York, NY (2002).

\end{thebibliography}\endgroup
\end{document}